\documentclass[aps,prl,reprint,superscriptaddress]{revtex4-1}

\usepackage{amsmath,bm}
\usepackage{graphicx}
\usepackage{dcolumn}
\usepackage{bm}
\usepackage{amssymb}
\usepackage{amsmath}
\usepackage{amsfonts}
\usepackage{epsf}
\usepackage{epsfig}
\usepackage{color}
\usepackage{epstopdf}

\usepackage{microtype} % <- makes text looks better just for nothing :D
\newcommand{\Fref}[1] {Fig. \ref{#1}}

\begin{document}
\title{
\large
Chiral photoelectron angular distributions \\ from ionization of achiral atomic and molecular species}

\author{Andreas~Pier}
\author{Kilian~Fehre}
\email{fehre@atom.uni-frankfurt.de}
\author{Sven~Grundmann}
\email{grundmann@atom.uni-frankfurt.de}
\author{Isabel~Vela-Perez}
\author{Nico~Strenger}
\author{Max~Kircher}
\author{Dimitrios~Tsitsonis}
\affiliation{Institut f\"ur Kernphysik, Goethe-Universit\"at, Max-von-Laue-Strasse 1, 60438 Frankfurt, Germany \\}
\author{Joshua~B.~Williams}
\affiliation{Department of Physics, University of Nevada, Reno, NV 89557, USA \\}
\author{Arne~Senftleben}
\affiliation{Institut f\"ur Physik und CINSaT, Universit\"at Kassel, Heinrich-Plett-Str. 40, 34132 Kassel, Germany \\}
\author{Thomas~Baumert}
\affiliation{Institut f\"ur Physik und CINSaT, Universit\"at Kassel, Heinrich-Plett-Str. 40, 34132 Kassel, Germany \\}
\author{Markus~S.~Sch\"offler}
\affiliation{Institut f\"ur Kernphysik, Goethe-Universit\"at, Max-von-Laue-Strasse 1, 60438 Frankfurt, Germany \\}
\author{Philipp~V.~Demekhin}
\affiliation{Institut f\"ur Physik und CINSaT, Universit\"at Kassel, Heinrich-Plett-Str. 40, 34132 Kassel, Germany \\}
\author{Florian~Trinter}
\affiliation{FS-PETRA-S, Deutsches Elektronen-Synchrotron (DESY), Notkestrasse 85, 22607 Hamburg, Germany}
\affiliation{Molecular Physics, Fritz-Haber-Institut der Max-Planck-Gesellschaft, Faradayweg 4-6, 14195 Berlin, Germany}
\author{Till~Jahnke}
\author{Reinhard~D\"orner}
\email{doerner@atom.uni-frankfurt.de}
\affiliation{Institut f\"ur Kernphysik, Goethe-Universit\"at, Max-von-Laue-Strasse 1, 60438 Frankfurt, Germany \\}

\date{\today}

\begin{abstract}
We show that the combination of two achiral components -- atomic or molecular target plus a circularly polarized photon -- can yield chirally structured photoelectron angular distributions.
For photoionization of CO, the angular distribution of carbon K-shell photoelectrons is chiral when the molecular axis is neither perpendicular nor (anti-)parallel to the light propagation axis.
In photo-double-ionization of He, the distribution of one electron is chiral, if the other electron is oriented like the molecular axis in the former case and if the electrons are distinguishable by their energy.
%These chiral emission patterns can be understood within the dipole approximation, 
%although a circularly polarized photon is not chiral in this approximation.
In both scenarios, the circularly polarized photon defines a plane with a sense of rotation and an additional  axis is defined by the CO molecule or one electron.
This is sufficient to establish an unambiguous coordinate frame of well-defined handedness.
To produce a chirally structured electron angular distribution, such a coordinate frame is necessary, but not sufficient.
We show that additional electron--electron interaction or scattering processes are needed to create the chiral angular distribution.

\end{abstract}
%\keywords{}
\maketitle
\section{Introduction}
An object is chiral if it cannot be brought to superposition with its mirror image by rotation or translation.
Often chirality is used in the context of molecular structures.
However, by definition the concept can be applied to any three-dimensional object.
Such an object can be a single particle wave function or its square modulus, i.e., a probability distribution in position or momentum space.
Here we discuss the chirality of photoelectron angular distributions (PADs), that are produced from the three-dimensional momentum vectors of photoelectrons.

Which ingredients are needed to enable the observation of a chiral PAD?
Experimentally, the emission pattern consists of individual photoionization events for each of which the direction of the photoelectron momentum vector is measured.
It may not come as a surprise that PADs arising from chiral molecules have a chiral structure in the molecular frame of reference, which can be obtained from multi-coincidence experiments (see e.g. \cite{Pitzer2013,Fehre2019}).
In this case, the chiral molecule itself provides a necessary prerequisite for a chirally structured PAD:
A coordinate system of well-defined handedness with respect to which the electron momentum vector can be measured.
We will demonstrate in this paper that the use of circularly polarized light and the detection of one non-coplanar vector in addition to the photoelectron is also sufficient to establish such a coordinate frame.

\begin{figure}
 \begin{center}
  \includegraphics[width=0.9\columnwidth]{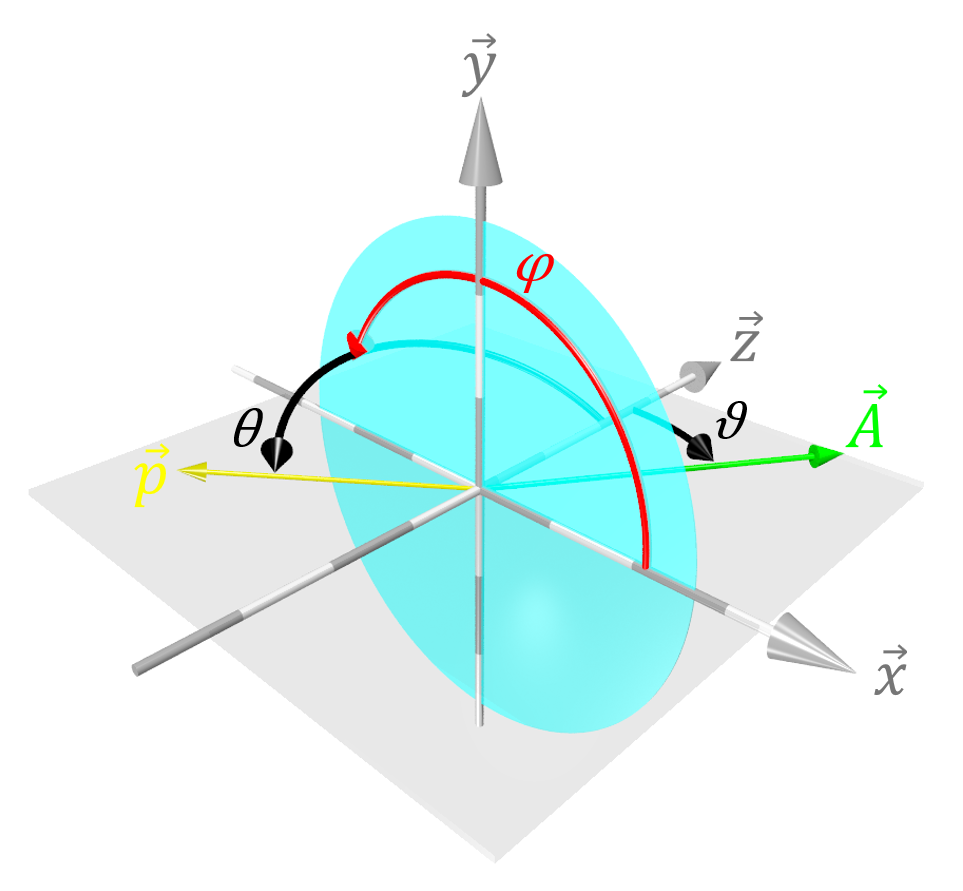}
 \end{center}
 \caption{Coordinate frame defined by a (prochiral) plane with sense of rotation, given by the circularly polarized light within the electric dipole approximation.
The vector $\vec{A}$ (green) is either the molecular axis in the CO case or the fast electron in photo-double-ionization of He.
The vector $\vec{p}$ (yellow) represents the electron, which is displayed in the angular distributions in the subsequent figures. 
Note that the polarization plane with its sense of rotation does not define the light propagation direction.
The positive direction of the z-axis is defined such that $\vec{z}$ and $\vec{A}$ lie on the same side of the polarization plane.
To display photoelectron angular distributions in this work, we define $\cos (\theta)= p_z / p$ and $\phi = \tan^{-1}(p_y/p_x)$, where $p_{x,y,z}$ are the respective vector components of the photoelectron momentum vector $\vec{p}$ of magnitude $p$.
To fix the orientation of $\vec{A}$ with respect to the prochiral plane, we define $\cos (\vartheta)= A_z / A$.
}
 \label{fig1}
\end{figure}

To introduce the coordinate frame in \Fref{fig1}, we first recap the properties of circularly polarized photons.
They are often considered a prototype of a chiral physical species because electric and magnetic field vectors describe a spiral in three-dimensional space.
For atomic and molecular photoionization, however, this chiral character is in many cases irrelevant as the pitch of the spiral is orders of magnitude larger than the object to be ionized \cite{Ordonez2018a}.
Thus, the spatial anisotropy of the light's vector potential, which is driving the ionization process, can often be neglected and light-matter interaction can be described evoking the dipole approximation.
Within this approximation, circularly polarized light is not chiral and the propagation direction is only a line, but not a vector (along the z-axis in \Fref{fig1}).
The circularly polarized light defines the polarization plane and a sense of rotation within that plane, but not the direction from which this plane has to be viewed.
Such a planar object is achiral in a three-dimensional world as its mirror image can be brought to superposition by a $180^\circ$ rotation around any axis in this plane.
In chemistry, such a planar object is referred to as \textit{prochiral}.
Note that the notation of left (LCP) and right circularly polarized (RCP) light does entail a direction of the $\vec{k}$ vector of the light, e.g. in the optical definition light is called RCP when the electric field vector rotates clockwise when looking towards the source.
Notorious confusion in the nomenclature  makes it necessary to state explicitly if one refers to looking with or against the $\vec{k}$ vector.
This makes it obvious that without reference to such a vector the meaning of RCP and LCP cannot be defined.
This is equivalent to saying that the sign of the magnetic quantum number entails the definition of a definite direction.
Nevertheless, within the dipole approximation this direction is not part of the interaction operator (see appendix of ~\cite{Ordonez2018a}).
For the remainder of this paper we implicitly assume the dipole approximation to be valid and ignore all possible effects resulting from magnetic or higher order transitions.   

Eventually adding the vector $\vec{A}$, which is not coplanar and not normal to the polarization plane, is sufficient to establish the handed coordinate frame in \Fref{fig1}.
%We choose the z-axis of the coordinate system to point in the direction of the $\vec{k}$ vector of the light. %Problem with the helium data, 60 and 120 is the same then
From the two antiparallel normal vectors $\vec{N}_{1,2}$ to the polarization plane we choose the one for which $\vec{A} \cdot \vec{N_{i}} > 0$ as the z-axis of the coordinate frame.
We choose the x-axis to be in the direction of the projection of $\vec{A}$ onto the polarization plane and the y-axis to be perpendicular to x and z with the positive direction being $90^\circ$ in the sense of rotation of the polarization vector. 
In the next section, we discuss an example in which $\vec{A}$ is given by the momentum vector of the faster of the two electrons in photo-double-ionization (PDI) of He. 
Afterwards the vector $\vec{A}$ is given by the molecular axis of a CO molecule pointing from Carbon towards Oxygen.

In our experiments on CO and He photoionization, we employed a COLTRIMS (Cold Target Recoil Ion Momentum Spectroscopy) reaction microscope \cite{Dorner2000, Ullrich2003, Jahnke2004a} and intersected a supersonic jet of the respective target gas with a synchrotron beam of circularly polarized photons from beamline P04 at $\text{PETRA III}$ (DESY, Hamburg \cite{Viefhaus2013}).
The reaction fragments from the interaction region were guided by electric and magnetic fields towards two time- and position-sensitive detectors \cite{Jagutzki2002,Jagutzki2002a}.
We detect all the reaction fragments in coincidence and calculate their three-dimensional momentum vectors from the times-of-flight and positions-of-impact.
The calculations supporting the CO experiment were performed within the dipole approximation by the stationary Single Center method \cite{Demekhin2011,Demekhin2007,Galitskiy2015}.

\section{Chiral electron emission in double ionization of He}
\label{seche}
In this section, we show that chiral electron emission patterns can be produced by one-photon double ionization (PDI) of a helium atom, i.e., a perfectly spherical symmetric initial state:
\begin{eqnarray}
h\nu _\text{circ} + He \rightarrow He^{2+} + 2e^- ~.
\end{eqnarray}
This process has been much studied in the past for linearly and circularly polarized photons (see \cite{Briggs00jpba,Doerner03rpa} for reviews).
Double ionization proceeds via three different mechanisms: knock-out, shake-off, and quasi-free. The latter is dipole-forbidden and contributes only at high photon energies \cite{Amusia75,Schoffler2013,Grundmann2018}.
In the former two mechanisms, the absorption of the photon leads to ejection of one electron and the second electron is then either knocked-out by a binary-type collision or is shaken-off due to the sudden change of the binding potential \cite{Schneider02prl,Kheifets01jpb,Knapp2002}.
The excess energy of the photon above the double ionization potential of helium (79 eV) is shared among the two electrons.
This energy sharing varies from being slightly enhanced at equal energy close to threshold to becoming more and more asymmetric with increasing photon energy.
The two electrons then become distinguishable by their energy.
Already in 1992, Berakdar and Klar \cite{Berakdar92} predicted that the process shows a circular dichroism, which has been confirmed experimentally in several works \cite{Viefhaus96prl,Mergel98prl,Kheifets98prl,Soejima99prl,Collins02pra,Achler01jpb,Knapp2005a}. 
In these experiments, mostly both electrons are kept in the polarization plane and the fully differential cross section is plotted as a function of the angle between the two electrons in that plane.
While this distribution inverts upon changing the polarization of the light, the emission distribution is in this coplanar geometry not chiral.
If, however, one selects one electron to be emitted out of the polarization plane, but not parallel to the light propagation, then this electron momentum vector can serve as $\vec{A}$ in \Fref{fig1}.

\begin{figure}[t]
 \begin{center}
  \includegraphics[width=1.0\columnwidth]{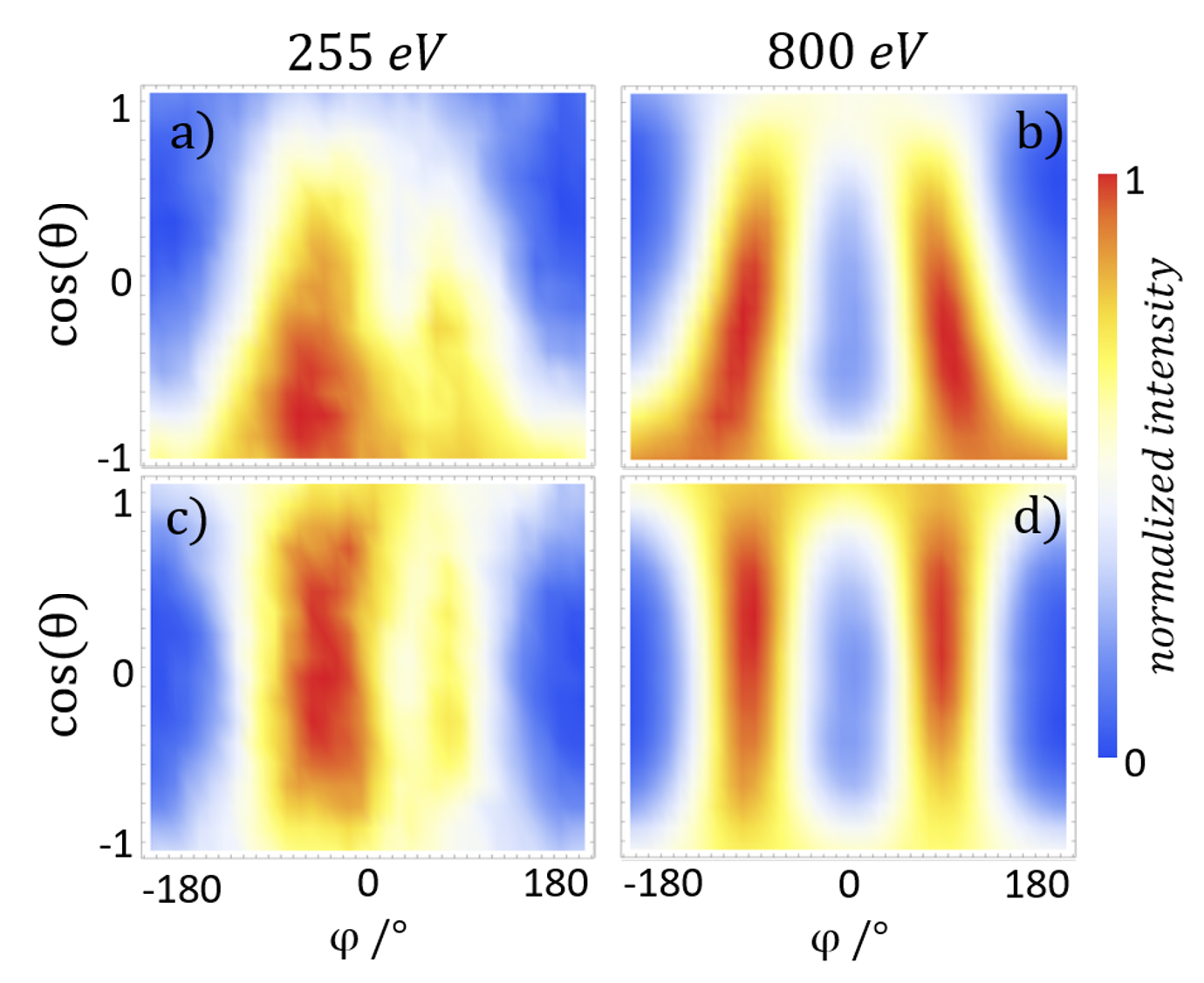}
 \caption{Angular distributions of the slow electron obtained from one-photon double ionization of the He by circularly polarized light at 255 eV (left column) and 800 eV (right column) photon energy, displayed in a coordinate frame as defined in \Fref{fig1}. 
The fast electron ($161 \pm 15$ eV for 255 eV photons and $686 \pm 35$ eV for 800 eV photons) defines $\vec{A}$ and encloses an angle with the z-axis of $60\pm20^\circ$ [(a) and (b)] or is held fix in the polarization plane [(c) and (d)].
In case of 255 eV photon energy, the angular distribution has an apparent chiral structure, but for 800 eV photon energy, electron--electron interaction is too weak and the chiral structure disappears, despite choosing a coordinate system of well-defined handedness in panel (b). }
 \label{figHe}
 \end{center}
\end{figure}

Figure \ref{figHe} shows photoelectron angular distributions for the slower electron of helium PDI with 255 eV (left column) and 800 eV (right column) circularly polarized photons, where $\cos (\theta)= p_z / p$ and $\phi = \tan^{-1}(p_y/p_x)$.
$p_{x,y,z}$ are the respective vector components of the photoelectron momentum vector $\vec{p}$ of magnitude $p$.
The figure shows a subset of the data from the same experimental session as \cite{Grundmann2018}, where experimental details can be found.
In panels (a) and (b), fast electrons with momentum vectors that have an angle of $60^\circ$ with the z-axis are selected.
%For panels (a) and (b) the fast electrons are emitted in the forward hemisphere of the light propagation and in the backward hemisphere for panels (e) and (f).
%Note that with this choice, the systems in the first and the last columns of \Fref{figHe} have opposed handedness and the shown PADs are mirror images of each other.
%For 255 eV photon energy, these mirror images are chirally structured as they cannot be superimposed by translation and rotation.
For 255 eV photon energy, the PAD in panel (a) is chirally structured as it cannot be superimposed by translation and rotation with its mirror image.
Similar to the case of molecular photoionization discussed in the next section, the effect does not follow from the definition of the coordinate frame alone.
It occurs by the joint action of the achiral photon and electron--electron interaction, which is also per se achiral.
The circularly polarized photon imprints a phase gradient in the electron wave as a function of the angle around the light propagation, which by electron--electron interaction leads to a chiral shape of the amplitude in three dimensional momentum space.
In other words, the circular dichroism \cite{Berakdar92} combined with an anisotropy in the electron mutual angle \cite{Knapp2005} forms a chiral object.
For 800 eV on the other hand, this circular dichroism is significantly lowered as electron--electron interaction becomes weaker with rising photon energy.
Thus, the PAD in Fig. \ref{figHe}(b) does not show chiral properties.
For Fig. \ref{figHe}(c) and \ref{figHe}(d), the fast electron's momentum vector is coplanar to the polarization plane.
As the coordinate frame has no well-defined handedness in this case, the resulting PADs cannot be chirally structured.

\section{Chiral electron emission from K-shell ionization of CO}
\label{secco}
Photoelectron angular distributions from fixed-in-space molecules are richly structured.
For emission from an inner-shell orbital of $s$ character, this structure is purely the result of multiple scattering of the outgoing photoelectron wave at the molecular potential.
The photoelectron wave "illuminates the molecule from within" \cite{Landers01prl}, creating a photoelectron diffraction pattern.
In the present case, we study emission of an electron from the carbon K-shell in CO.
The K-hole relaxes by Auger decay creating $CO^{2+}$, which can dissociate into $C^+ + O^+$ \cite{Weber03prl}.
We measure the photoelectron energy and angle, the kinetic energy release (KER), and the direction of the two ionic fragments in coincidence.
For high-lying electronic states leading to a KER above 10.2 eV, this decay is much faster than the rotational motion of the molecule \cite{Weber01jpb}, allowing us to infer the molecular orientation at the instant of photoabsorption from the direction of the measured fragment ions. 
CO was one of the first molecules for which molecular-frame photoelectron angular distributions have been reported for both linearly and circularly polarized light \cite{Schoenhense96,Ito00prl,Adachi93prl,Heiser97prl,Cherepkov00jpb,Landers01prl,Jahnke02prl,Weber01jpb}.
In all previous studies using circularly polarized light, emission patterns have been reported just for molecular orientations in the polarization plane. 
In this case, the PADs are symmetric upon reflection at the polarization plane and thus not a chiral object within the dipole approximation.
As it can be seen in \Fref{fig1} for molecules aligned in the polarization plane, not even a coordinate frame of well-defined handedness is established within the dipole approximation.
This is different for the choice of CO enclosing an angle of $70^\circ$ with the z-axis, where a handed coordinate frame is uniquely established.

\begin{figure*}
 \begin{center}
  \includegraphics[width=1.47\columnwidth]{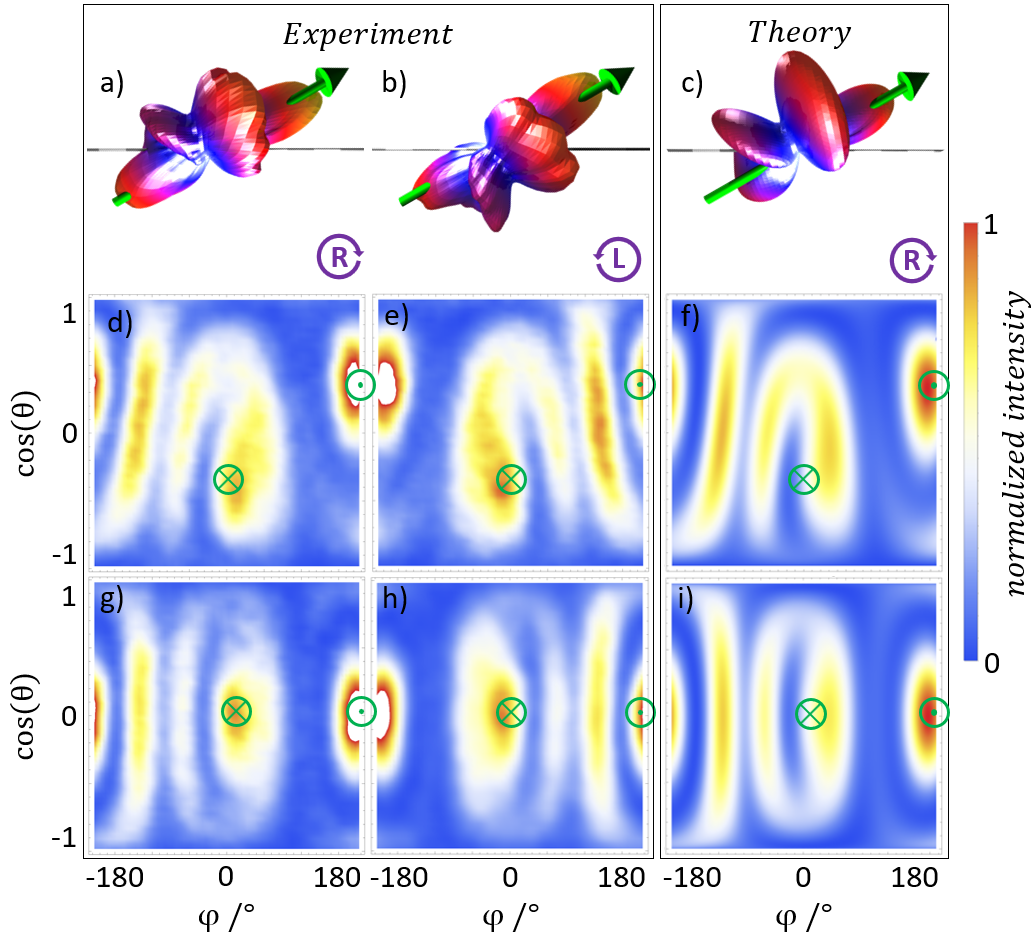}
 \end{center}
   \caption{Photoelectron angular distributions of the carbon K-shell electron emission of CO, obtained from circularly polarized light at 310 eV, presented in the coordinate frame as defined in \Fref{fig1}. The molecular axis of the CO molecule, pointing from the carbon atom to the oxygen atom is defined as vector $\vec{A}$ and is selected to enclose an angle with the z-axis of $70 \pm 4^\circ$ (a)--(f) or to lie in the polarization plane (g)--(i). Vector $\vec{A}$ is displayed as the green arrow in (a)--(c) and the green markers in (d)-(i).
%For the PADs shown in (a)-(d) [(e) and (f)], the enclosed angles between the molecular axis $\vec{A}$ and the polarization plane are selected to be $20^\circ$ [$0^\circ$].
%Panels (a)-(d) display the PADs, for the case when the CO molecule (vector $\vec{A}$)  and the polarization plane enclose an angle of $20^\circ$, while panels (e) and (f) show the PADs when the molecule lies in the polarization plane.
In the first row [(a)--(c)] the intensity of the angular distribution is encoded by the distance of the surface to the origin. The panels in the second row [(d)--(f)] show the same data, but the normalized intensity is represented by a colormap.
The two columns on the left display experimental data for right-handed circularly  [(a),(d), and (g)] and left-handed circularly polarized light [(b),(e), and (h)], whereas the panels in the right column [(c),(f), and (i)] show calculations for right-handed circular polarization. Comparisons between figures (a)--(f) with figures (g)--(i) show that chiral structures are apparent if the molecules lie outside the polarization plane, but vanish if they lie inside.}
\label{figCO}
\end{figure*}

In \Fref{figCO}, we show K-shell PADs in this coordinate frame at 310 eV photon energy.
The panels (a), (d), and (g) [(b), (e), and (h)] show experimental data for right-handed [left-handed] circular polarization, and panels (c), (f), and (i) display calculations for right-handed circular polarization.
Figures \ref{figCO}(a) and \ref{figCO}(d) [(b) and (e), (c) and (f)] show the same result in different representations where CO is oriented $70^\circ$ to the z-axis.
In panels (a), (b), and (c) the angular variation of the electron yield is encoded in the distance of the surface from the origin.
Panels (d), (e), and (f) show the same PADs, but in three-dimensional spherical polar coordinates where the yield is color-coded.
The shape of these PADs is clearly a chiral structure.
Inversion of the light polarization %, i.e. comparing experiment and theory,
changes the PAD to its mirror image, and agreement between theory and experiment is excellent.
The calculations are performed within the dipole approximation, highlighting that the chirality of the PAD is created by multiple scattering and not by nondipole contributions to the light-matter interaction.
In Figs. \ref{figCO}(g), \ref{figCO}(h), and \ref{figCO}(i), we show the PADs for experiment and theory when CO is coplanar to the polarization plane.
In this case, the coordinate frame has no well-defined handedness and the resulting PADs are not chirally structured.

Like in the former case of He, the ability of our experimental arrangement to define a handed coordinate frame, does not automatically, i.e., by definition, entail that the electron distribution has to be chiral.
This is made by the physical effect of multiple scattering.
The circularly polarized light encodes its sense of rotation in the phase of the photoelectron wave emerging from the spherically symmetric K-shell.
It is the scattering of this complex-valued wave at the $CO^+$ potential which translates the angular dependent phase into an amplitude which then can be measured.
It is the joint action of two achiral ingredients, an achiral circular photon and the scattering in a potential of a linear molecule, which taken together give rise to the chiral electron emission pattern. 

%\textcolor{red}{Andi, hier kommt dann Beitrag zu high-energy CO rein. Scattering gets less with higher energy, no chiral structure!? Vielleicht doch chiral?}

\section{Connection to PECD}
Photoelectron circular dichroism (PECD), as the word is used in most works today, refers to a forward/backward asymmetry of electron emission with respect to the light propagation direction that inverts upon inversion of the light helicity.
The effect occurs for one- or multi-photon ionization of randomly oriented chiral molecules, i.e., without the need for any coincidence detection \cite{Ritchie1976,Bowering2001,Nahon2006,Nahon2010,Nahon2015a,Lux2012a,Lux2015a,Lux2015b,Tia2017,Rozen2019,Fehre2019a}.
For PECD, the chiral structure of the molecule acts as a gearbox which translates the rotation of the electric field vector into a linear forward or backward motion of the emitted electron.

Creating a chiral PAD is not enough to create a PECD according to this definition.
Changing the helicity of the light leads to a mirror-symmetrical PAD [see e.g. \Fref{figCO}(a) and \Fref{figCO}(b)], but it does not reverse the observed forward/backward asymmetry along the light propagation direction [encoded in $\cos(\theta)$ in \Fref{fig1}].
However, our coincident detection scheme allows to unveil an apparent PECD under certain geometry configurations.

\begin{figure}
 \begin{center}
  \includegraphics[width=1.0\columnwidth]{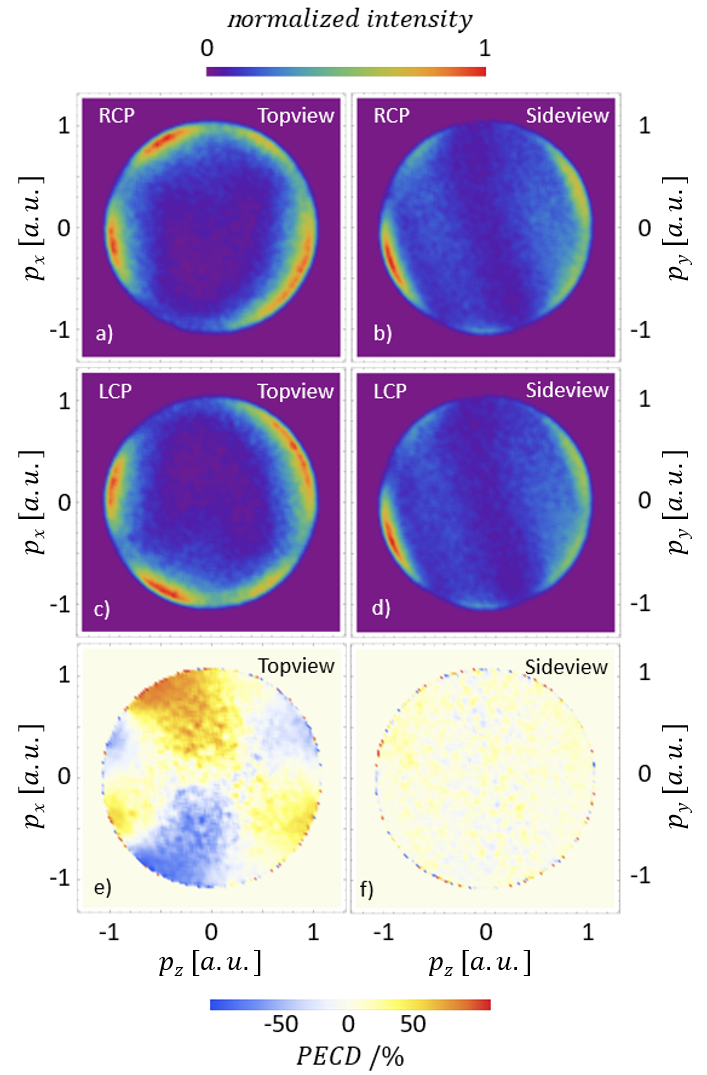}
 \end{center}
 \caption{Electron momentum distributions for photoionization of CO by 310 eV circularly polarized light. Panels (a) and (c) [(b) and (d)] show the two-dimensional projections of the three-dimensional photoelectron momenta onto the x-z plane [y-z plane] as defined by the coordinate frame in \Fref{fig1}. The first row, panels (a) and (b), shows these projections for a right circularly polarized (RCP) photon, whereas the second row displays the distributions for the case of left circularly polarized (LCP) light. The normalized differences of the momentum distributions for the two polarization states are displayed in (e) for the topview (projection onto the x-z plane) and in (f) for the sideview (projection onto the y-z plane). For the topview (left column) apparent PECD is identifiable, while the sideview (right column) shows no signs of PECD.  }
 \label{fig4}
\end{figure}

In \Fref{fig4}, we project the three-dimensional electron momentum vectors from CO photoionization at 310 eV photon energy into a two-dimensional coordinate system, which contains the light propagation direction (z-axis).
The second axis is either the x-axis (topview, left column) or the y-axis (sideview, right column) as defined in \Fref{fig1}.
Note that in the sideview case, the molecular axis appears to be parallel to the light propagation in the chosen two-dimensional coordinate system. 
The first row in \Fref{fig4} shows the respective momentum distribution with right-handed circular polarization (RCP) and the middle row with left-handed circular polarization (LCP).
In the the topview geometry, switching the helicity of the light flips the upper and lower half of the momentum distribution.
The third row shows the two-dimensional PECD maps, generated from normalized differences of RCP and LCP momentum distributions.
While the sideview PECD map displays no significant structure, the topview shows that the coincidence detection allows to observe a strong apparent PECD.
For the upper (lower) half alone, there is a pronounced forward/backward asymmetry in panel (e).
However, the total amount of forward-emitted electrons does, even with coincidence detection, not change with the light helicity.
Thus, integration along the x-axis in the topview PECD map spoils the effect and underlines that real PECD 
requires the gearbox effect of a chiral molecular potential.

\section{Conclusion}
We have shown that measuring a photoelectron in coincidence with another particle, such as a second electron or a fragment ion from CO, allows to measure a chiral electron emission pattern, i.e., a density distribution in momentum space which is chiral.
This distribution is chiral despite the fact that the initial state --  the circularly polarized light in the dipole approximation and the atom or linear molecule -- consists of only achiral ingredients. This raises the question if achiral ingredients could also cause a measurable PECD~\cite{Ordonez2019}, which is of great practical relevance for experiments that try to use PECD as probe for molecular chirality and enantiomeric excess. One type of these experiments rely on resonance enhanced multiphoton ionization (REMPI) where the ionization step is preceded by photoexcitation. Such excitation does select certain molecular alignments and hence the ionization occurs out of a potentially aligned ensemble. The alignment could establish the vector $\vec{A}$ in \Fref{fig1}. The examples we have shown here indicate that creating a one-dimensional alignment or even orientation can only produce what we call an apparent PECD, but the effect vanishes after integration. A second class of experiments where this could be important are pump--probe experiments where a chiral molecule is dissociated by a pump pulse and one attempts to use PECD as probe for the chirality of the fragments. In this scenario, the dissociation axis can establish an additional axis. Also in this case, our results show that no PECD can be produced in such a dissociation unless the fragments themselves are chiral. In conclusion, chiral PADs and PECD do not entail each other.     
\section{Acknowledgments}
\begin{acknowledgments}
We acknowledge DESY (Hamburg, Germany), a member of the Helmholtz Association HGF, for the provision of experimental facilities.
Parts of this research were carried out at PETRA III and we would like to thank the staff for excellent support during the beam time.
Parts of this research were funded by the Deutsche Forschungsgemeinschaft (DFG, German Research Foundation), Projektnummer 328961117, SFB 1319 ELCH.
We acknowledge support by BMBF.
K.F. acknowledges support by the German National Merit Foundation.
M.S. acknowledges financial support by the Adolf Messer foundation.
\end{acknowledgments}

\end{document}